%% file: ms.tex
\documentclass[pdflatex,a4paper,twocolumn]{article}
\usepackage{PRIMEarxiv}
\usepackage{cite}
\usepackage{amsmath,amssymb,amsfonts}
\usepackage{physics}
\usepackage{algorithm}
\usepackage{algpseudocode}
\usepackage{graphicx}
\usepackage{textcomp}
\usepackage{float}
\usepackage{gensymb}
\usepackage{booktabs}
\usepackage{comment}
\usepackage{color}
\usepackage{fancyhdr}

\makeatletter
\let\MYcaption\@makecaption
\makeatother
\usepackage[labelformat=simple]{subcaption}

\makeatletter
\let\@makecaption\MYcaption
\makeatother

\usepackage[binary-units]{siunitx}[=2021-04-09]
\begin{document}
\title{Experimental Demonstration of Delay-Bounded Wireless Network Based on Precise Time Synchronization\footnote{This work was supported in part by the New Energy and Industrial Technology Development Organization (NEDO) under Grant JPNP20017, in part by the Japan Science and Technology Agency through the Core Research for Evolutionary Science and Technology (CREST) Project under Grant JPMJCR17N2, and in part by the Japan Society for the Promotion of Science through the Grants-in-Aid for Scientific Research (A) under Grant JP20H00233.}}

\author{%
Haruaki Tanaka\footnote{Department of Information Physics and Computing, Graduate School of Information Science and Technology, The University of Tokyo, Tokyo 113-8656, Japan}\and
Yusuke Yamasaki\footnotemark[2]\and
Satoshi Yasuda\footnote{Space-Time Standards Laboratory, Electromagnetic Standards Research Center, Radio Research Institute, National Institute of Information and Communications Technology, Koganei 184-8795, Japan}\and
Nobuyasu Shiga\footnotemark[3]\and
Kenichi Takizawa\footnote{Sustainable ICT System Laboratory, Resilient ICT Research Center, Network Research Institute, National Institute of Information and Communications Technology, Sendai 980-0812, Japan}\and
Nicolas Chauvet\footnotemark[2]\and
Ryoichi Horisaki\footnotemark[2]\and
Makoto Naruse\footnotemark[2]}
\date{}

\maketitle

\begin{abstract}
  Low latency and reliable information transfer are highly demanded in fifth generation (5G) and beyond 5G wireless communications.
  A novel delay-bounded wireless media access control (MAC) protocol called Carrier Sense Multiple Access with Arbitration Point (CSMA/AP) was established to strictly ensure the upper boundary of communication delay.
  CSMA/AP enables collision-free and delay-bounded communications with a simple arbitration mechanism exploiting the precise time synchronization achieved by Wireless Two-Way Interferometry (Wi-Wi).
  Experimental demonstration and proving the feasibility in wireless environments are among the most critical steps before any further discussion of CSMA/AP and extension to various applications can take in place.
  In this work described in this paper, we experimentally demonstrated the fundamental principles of CSMA/AP by constructing a star-topology wireless network using software-defined radio terminals combined with precise time synchronization devices.
  We show that CSMA/AP was successfully operated, even with dynamic changes of the spatial position of the terminal or the capability to accommodate mobility, thanks to the real-time adaption to the dynamically changing environment by Wi-Wi.
  We also experimentally confirmed that the proposed CSMA/AP principle cannot be executed without Wi-Wi, which validates the importance of precise time synchronization.
  This study paves the way toward realizing delay-bounded wireless communications for future low-latency and highly reliable critical applications.
  {\flushleft{\textbf{keywords: } Bounded delay; CSMA; MAC protocol; precise time synchronization; wireless two-way interferometry; wireless network}}
\end{abstract}

\section{Introduction} \label{sec:introduction}

\subsection{Background}

The explosively increasing importance of wireless communications requires high-speed operation and reliability in various applications and industries~\cite{adhikari2022comstd}.
Since the release of a recommendation from ITU-R about International Mobile Telecommunications for 2020 and beyond (IMT-2020)~\cite{ITU-RM.2083-0}, many researchers have worked on key technologies for Fifth Generation (5G) mobile communications~\cite{gai2021icbaie} and are discussing features of the coming beyond 5G or Sixth Generation (6G) wireless communications~\cite{alsabah2021access,ray2021sysarc}.
IMT-2020~\cite{ITU-RM.2083-0} mentioned three usage scenarios, and many use cases under discussion for 6G communications are also based on these studies~\cite{saad2020network}.

The Ultra-Reliable Low Latency Communication (URLLC) usage scenario in IMT-2020~\cite{ITU-RM.2083-0} is about applications of radio communications that have stringent communication delay and reliability requirements, e.g., healthcare, transport and automotive, entertainment, and manufacturing~\cite{lema2017access}.
From the perspective of reliable communications, a strictly bounded time delay is crucial for real-time systems, especially hard real-time systems imposing a rigid deadline or predictable real-time performance where the worst latency is predicted before communication takes place.
However, contemporary wireless communication methods require complex arbitration mechanisms to deal with collisions (i.e., interference) caused by multiple simultaneous access.
For example, Carrier Sense Multiple Access with Collision Avoidance (CSMA/CA), a widely used arbitration method for wireless local area networks, relies on the random back-off algorithm to avoid collisions.
CSMA/CA shows good performance when the traffic load is low; however, the wait time increases significantly when the traffic load is high.
Moreover, due to their stochastic behavior, communication delays over CSMA/CA cannot be deterministically predicted even in a low-traffic regime.

Deterministic and Synchronous Multi-channel Extension (DSME), Low Latency and Deterministic Network (LLDN), and Time-Slotted Channel Hopping (TSCH)~\cite{ramonet2020icact,choudhury2020access}, defined in the IEEE 802.15.4 standard, are MAC behaviors guaranteeing low and deterministic latency.
They all involve approaches based on Time Division Multiple Access (TDMA), and devices synchronize their behaviors by using periodic beacon signals.
However, fixed dedicated time slots for a particular terminal would be wasted when the terminal has nothing to transmit, reducing the overall throughput.
Other arbitration methods with guarantees of maximum delay have also been proposed~\cite{ergen2006tmc,sahoo2010tmc,zhang2012icarcv,farag2018jsen}.
However, they have difficulties with scalability, low-traffic performance, and equality among terminals.

\begin{figure}[tb]
  \centering
  \begin{minipage}{70mm}
    \centering
    \includegraphics[width=70mm]{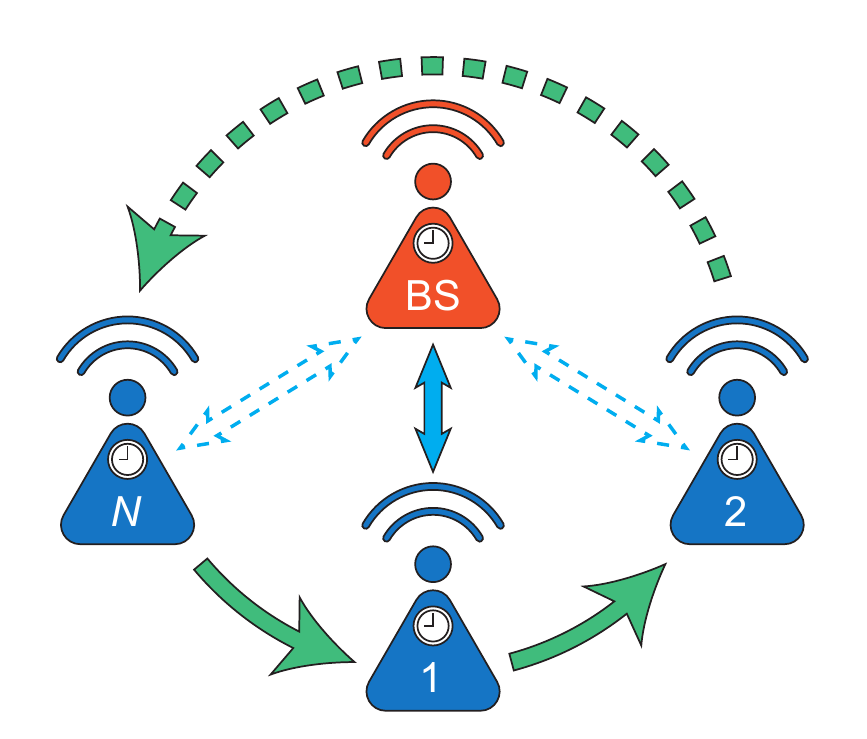}
    \subcaption{}
    \label{fig:introduction:concept:csma_ap}
  \end{minipage}
  \begin{minipage}{70mm}
    \centering
    \includegraphics[width=70mm]{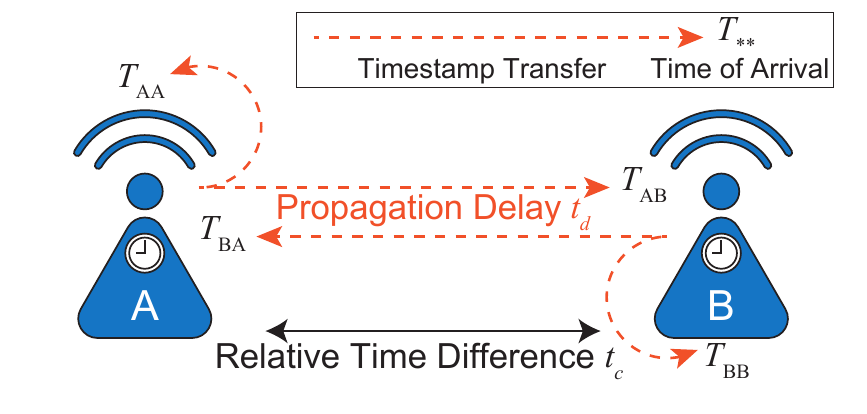}
    \subcaption{}
    \label{fig:introduction:concept:wi-wi}
  \end{minipage}
    \caption{Design principle of CSMA/AP.
    \subref{fig:introduction:concept:csma_ap}~Schematic diagram of CSMA/AP.
    Each terminal acquires periodical transmission opportunities, recognizing them with its precisely synchronized clock.
    This mechanism limits the maximum communication delay.
    \subref{fig:introduction:concept:wi-wi}~Schematic diagram of Wireless Two-Way Interferometry (Wi-Wi).
    The relative time difference of two clocks at each site and the distance between them are calculated by exchanging timestamps $T_\mathrm{AA}, T_\mathrm{AB}, T_\mathrm{BA}$ and $T_\mathrm{BB}$, and measuring carrier phases.
    }
  \label{fig:introduction:concept}
\end{figure}

Concerning the above backgrounds, we consider that time synchronization is a promising feature in providing highly reliable wireless communications.
Indeed, Yamasaki \textit{et al}.~\cite{yamasaki2021access} proposed a novel MAC method named Carrier Sense Multiple Access with Arbitration Point (CSMA/AP), to accomplish strictly delay-bounded wireless communication by exploiting the feature that all devices are precisely time-synchronized.
In CSMA/AP, as schematically illustrated in Fig.~\ref{fig:introduction:concept:csma_ap}, communication opportunities are specified by Arbitration Points (AP), which are periodically assigned for all devices.
Hence, CSMA/AP provides all terminals with collision-free and equal transmission opportunities.
It should be emphasized that CSMA/AP guarantees the upper bound of communication delay and entirely relies on precise time synchronization.

Precise time synchronization itself is required by telecommunication technologies like Massive Multiple-Input Multiple-Output (MIMO)~\cite{larsson2014mcom,ammar2022csto} and technologies in other industries, e.g., factory automation, power delivery networks, vehicular communications, and wireless live audio/video production~\cite{mahmood2019mcom}.
Several time synchronization technologies are established and used for mobile communications.
One widely used technology is the Global Navigation Satellite System (GNSS)~\cite{maroti2004acm}.
However, its synchronization performance degrades depending on the environment being used.
In particular, satellite information is unavailable in typical indoor environments.
The Precision Time Protocol (PTP)~\cite{ieee-1588-2019} is another way to achieve precise time synchronization.
However, its precision and applicability are limited due to its inability to manage dynamically changing uncertain network structures, especially on short time scales.

In light of this background and context, Shiga \textit{et al}.~\cite{shiga2017comex} have developed a time synchronization technology called Wireless Two-Way Interferometry (Wi-Wi for short), depicted in Fig.~\ref{fig:introduction:concept:wi-wi}.
Wi-Wi is the most feasible enabler for providing precise time synchronization among many terminals, as required by CSMA/AP.
In Wi-Wi, the devices in the network exchange their time information with each other, allowing them to update their internal clocks autonomously and recognize relative spatial distance information.
That is, time- and space-synchronization are accomplished by Wi-Wi.
The principle of Wi-Wi is reviewed in Sec.~\ref{sec:preliminaries} below.
Ultimately, terminals with Wi-Wi are time-synchronized with picosecond-level precision.
It should be emphasized that Wi-Wi does not need expensive instruments; commercially available semiconductor devices and crystal oscillators are the underlying components of Wi-Wi~\cite{shiga2017comex}.
These latest advancements in time- and space-synchronization are the foundations of the present study regarding delay-bounded communications.

\subsection{Overview of This Paper}

The former study of CSMA/AP~\cite{yamasaki2021access}, however, was limited to theoretical and numerical demonstrations.
Although \cite{yamasaki2021access} also presented preliminary experimental studies based on wired connections, the effect of Wi-Wi-specific features on CSMA/AP has not been clarified.
However, it is expected that CSMA/AP will be applied the arbitration of communications among distributed mobile terminals.
Thus, experimental demonstrations and verifications are indispensable for examining the feasibility and the utility of time synchronization for delay-bounded communications in wireless environments.

Therefore, in this paper, we demonstrate the fundamental principles of CSMA/AP by implementing them in software-defined radio (SDR) systems operating in the \SI{2.4}{\giga\hertz} regime in combination with wireless time synchronization devices, called Wi-Wi modules, operating in the \SI{920}{\mega\hertz} communication band.
We demonstrate the successful operation of CSMA/AP in wireless environments, including its ability to manage devices that move around in the configured network.
We also experimentally demonstrate that CSMA/AP does not work without precise time synchronization.
Meanwhile, we experimentally characterized the precise time synchronization itself by redefining frequency stability as the stability of synchronization to the reference clock.
The main contributions of our study are summarized as follows.
\begin{itemize}
    \item We characterized the precise time synchronization of Wi-Wi by utilizing frequency stability measurement. Consequently, we found that long-term stability is a unique attribute of Wi-Wi (Sec.~\ref{sec:stability}).
    \item The fundamental principles of CSMA/AP were implemented in SDR systems combined with Wi-Wi modules (Sec.~\ref{sec:experiment:detail}).
    \item The sensitivity of the proper arbitration of CSMA/AP on SDR to the protocol-specific time parameters was inspected, and the operating range of these parameters was clarified (Sec.~\ref{sec:experiment:ap}).
    \item We demonstrated successful communication on CSMA/AP with a moving terminal, which is challenging in wired networks (Sec.~\ref{sec:experiment:mobility}).
    \item We clarified the necessity for the precise time synchronization by Wi-Wi, by comparing the operation of CSMA/AP with or without time synchronization (Sec.~\ref{sec:experiment:async}).
\end{itemize}

This paper is organized as follows.
Sec.~\ref{sec:preliminaries} reviews the fundamentals of Wi-Wi and CSMA/AP.
Sec.~\ref{sec:stability} demonstrates the synchronization ability of Wi-Wi on a variety of timescales.
Sec.~\ref{sec:experiment} describes experimental demonstrations of CSMA/AP and presents the results.
Sec.~\ref{sec:conclusion} concludes the paper.

\section{Preliminaries} \label{sec:preliminaries}

\subsection{Time Synchronization Technology: Wi-Wi} \label{sec:preliminaries:wi-wi}

Wireless Two-Way Interferometry (Wi-Wi)~\cite{shiga2017comex} is a technology that achieves precise time and frequency synchronization among clocks.
It employs techniques for standard time comparison between remote sites~\cite{hanson1989frequency} and synchronizes oscillators.
By sharing precisely synchronized clocks, the variation in distance between antennas can also be obtained.

The fundamentals of Wi-Wi come from Two-Way Satellite Time and Frequency Transfer (TWSTFT)~\cite{hanson1989frequency}, which was developed to compare standard time via geostationary satellite signals.
There are two technologies: Two-Way Time Transfer (TWTT)~\cite{hanson1989frequency} and Carrier-Phase Two-Way satellite frequency transfer (TWCP)~\cite{fujieda2012tuffc}.

We first review the principle of time synchronization using TWTT.
TWTT assumes that clocks exchange their time information via two-way communication through the same physical route.
$t_c$ is the time difference between the two clocks; in other words, a time point is observed by the clock B $t_c$ after the clock A observes the same time point.
Then, the time difference calculated when the terminal at A receives time information after the propagation delay $t_d$ is given as follows:
\begin{equation}
  t_\mathrm{A} = T_\mathrm{BA} - T_\mathrm{BB} = -t_c + t_d \label{eq:wi-wi:t_A}
\end{equation}
where $t_\mathrm{A}$ denotes the time difference calculated at A.
$T_\mathrm{BB}$ and $T_\mathrm{BA}$ are the time when time information from B arrives at B and A, respectively.
Also, the time difference is obtained at B in the same way as follows:
\begin{equation}
  t_\mathrm{B} = T_\mathrm{AB} - T_\mathrm{AA} = t_c + t_d \label{eq:wi-wi:t_B}
\end{equation}
where $t_\mathrm{B}$ denotes the time difference calculated at B.
$T_\mathrm{AA}$ and $T_\mathrm{AB}$ are when time information from A arrives at A and B, respectively.
Assuming that the single round trip time is sufficiently small, we can obtain the time difference $t_c$ and the propagation delay $t_d$ as follows:
\begin{align}
  t_c & = \frac{t_\mathrm{B} - t_\mathrm{A}}{2} \label{eq:wi-wi:t_c} \\
  t_d & = \frac{t_\mathrm{B} + t_\mathrm{A}}{2} \label{eq:wi-wi:t_d}.
\end{align}
Therefore, obtaining $t_c$ allows the synchronization of both clocks.

TWCP provides the time difference information in the same manner by using the carrier phase instead of timestamps.
We can rewrite $t_c$ by using the carrier phase information:
\begin{equation}
  t_c = \frac{\phi_c}{4\pi f_0} = \frac{\phi_\mathrm{B} - \phi_\mathrm{A} + 2\pi M_c}{4\pi f_0} \label{eq:wi-wi:phi_c}
\end{equation}
where $f_0$ is the carrier frequency, $\phi_c = \phi_\mathrm{B} - \phi_\mathrm{A} + 2\pi M_c$ is the phase difference between two terminals, and $M_c$ is an integer.
$\phi_\mathrm{A}$ and $\phi_\mathrm{B}$ are obtained from measured phases, similarly to \eqref{eq:wi-wi:t_A} and \eqref{eq:wi-wi:t_B}.
While $t_c$ cannot be determined by only TWCP due to the integer ambiguity $M_c$, the variation of $t_c$ can be tracked over time if the variation of $\phi_c$ between successive observations is less than $\pi$.

Based on the technique described above, Wi-Wi can synchronize terminals at remote sites via wireless communication.
Each terminal has to calculate quantities given by \eqref{eq:wi-wi:t_c}, \eqref{eq:wi-wi:t_d}, and \eqref{eq:wi-wi:phi_c} to get the actual time difference, meaning that one and a half roundtrip timestamp exchanges and carrier phase measurements are needed.

With precisely synchronized clocks, we can measure the distance between them as follows:
\begin{equation}
  \begin{split}
    l_d & := c \cdot t_d = - \frac{c}{4\pi f_0}(\phi_\mathrm{A} + \phi_\mathrm{B} + 2\pi M_c) \\
        & = - \left( \frac{\phi_\mathrm{A} + \phi_\mathrm{B}}{2\pi} + M_c\right) \frac{c}{2\pi f_0}
  \end{split} \label{eq:wi-wi:distance}
\end{equation}
where $l_d$ is the propagation distance between the two terminals, which is equal to the product of the speed of light $c$ and the propagation delay $t_d$~\cite{panta2019sam}.
There remains the integer ambiguity $M_c$; however, this can be resolved in the same manner as the measurement of $t_c$.
Also, the horizontal atmospheric propagation delay and, therefore, the water vapor distribution can be measured~\cite{yasuda2019rs}.

Yasuda \textit{et al}.~\cite{yasuda2019rs} have developed a compact electronic module implementing Wi-Wi principles, referred to as the Wi-Wi module.
When Wi-Wi modules are in a state called phase-lock, the built-in crystal oscillators are synchronized to the leader module's oscillator by PID feedback control and provide precise time-synchronized one pulse-per-second (PPS) signals and \SI{10}{\mega\hertz} signals.
The PPS signals that the Wi-Wi modules provide have synchronization errors that roughly follow a normal distribution with $3\sigma = \SI{400}{\nano\second}$~\cite{kameda2021icufn}.

\subsection{Delay-Bounded MAC Protocol: CSMA/AP} \label{sec:preliminaries:csma_ap}

Carrier Sense Multiple Access with Arbitration Point (CSMA/AP) is a contention-based media access control protocol for wireless networks, realizing collision-free arbitration by employing precise time synchronization~\cite{yamasaki2021access}.
Thanks to the collision-free property and the equality of transmission opportunities among terminals, CSMA/AP can guarantee an upper boundary of the communication delay.
Here, we briefly summarize the principle of CSMA/AP.

\begin{figure}[tb]
  \centering
  \includegraphics[width=70mm]{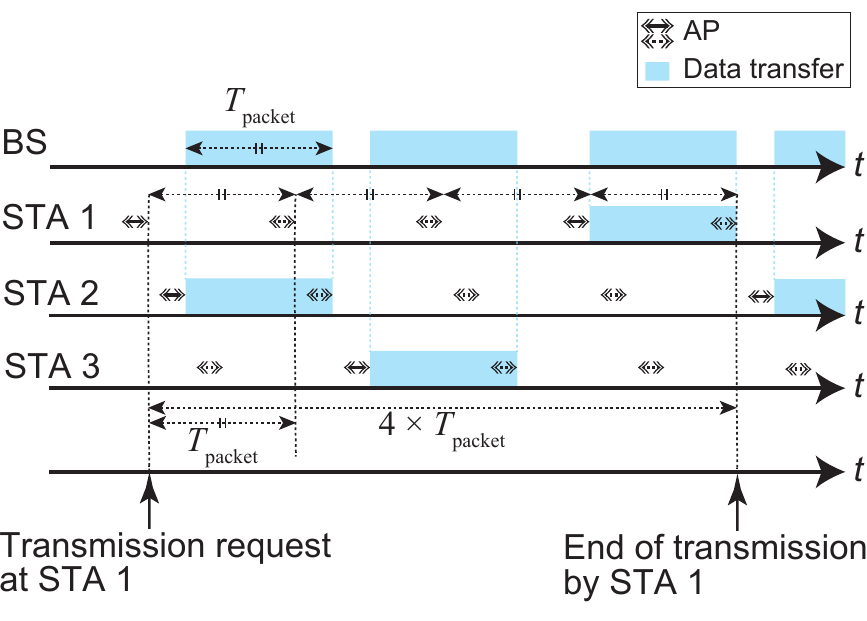}
  \caption{Example timing chart of APs and data transmissions controlled by CSMA/AP when the number of terminals is three.
  Network participants start packet transmission (denoted by rectangles) right after each AP (denoted by left-right two-headed arrows) if the others are not sending data.}
  \label{fig:csma_ap:timing}
\end{figure}

The core idea of CSMA/AP is the notion of an Arbitration Point (AP), a time duration periodically assigned and shared among all terminals.
Each terminal is periodically assigned to a completely distinct AP.
Furthermore, the timing of the AP for a terminal is autonomously detected thanks to precise time synchronization.
Fig.~\ref{fig:csma_ap:timing} is the timing chart of APs and data transmissions controlled by CSMA/AP.
When a terminal undergoes the time duration of an AP, it checks or senses the environment during this AP duration; namely, carrier sense is conducted.
If the terminal senses that the channels idle in this designated AP duration, it can start data transmission.
Otherwise, it must postpone the transmission until the next AP timing.
Based on the periodic assignment of APs to the terminals, the terminal can surely start data transmission no later than $(N + 1)$ rounds of AP duration when $N$ terminals are in the network, leading to strictly ensuring the upper bound of delay~\cite{yamasaki2021access}.

CSMA/AP has two variants; CSMA/AP-T and CSMA/AP-TS~\cite{yamasaki2021access}.
CSMA/AP-T utilizes only time synchronization and distributes APs in equally spaced time intervals.
Meanwhile, CSMA/AP-TS employs space localization in addition to time synchronization and optimizes assignment order and intervals between successive APs.
Remember that Wi-Wi allows capturing of inter-device distance information, not just time-difference information.
Therefore, the duration of AP can be configured in a much denser manner in the time domain by CSMA/AP-TS than by CSMA/AP-T.

\section{Wi-Wi Frequency Stability and Synchronization} \label{sec:stability}

Atomic clocks are more stable than ordinary crystal oscillators in terms of frequency stability.
However, what is critical for CSMA/AP is that all devices are synchronized.
That is, synchronization is more important than the stability of individual oscillators.
To quantitatively examine the impact of synchronization, here we evaluate the stability of synchronization using rubidium clocks and Wi-Wi modules.

We used the Allan deviation~\cite{kartaschoff1978academic} to evaluate the stability of synchronization.
Let $y$ denote the fractional frequency offset:
\begin{equation}
  y(t) = \frac{1}{2\pi\nu_0} \dv{\Phi(t)}{t}
\end{equation}
where $\Phi$ is the difference in instantaneous phase between the reference clock signal and the measurement target signal, and $\nu_0$ is the nominal frequency.
In practice, the time average over averaging time $\tau$:
\begin{equation}
  y(t; \tau) = \frac{1}{\tau} \int_t^{t+\tau} y(t') \dd{t}'
\end{equation}
is obtained, and the Allan deviation $\sigma_y(\tau)$ is defined as follows:
\begin{equation}
  \sigma_y(\tau) = \sqrt{\left< \pqty{y(t + \tau; \tau) - y(t; \tau)}^2 \right>}.
\end{equation}
The Allan deviation $\sigma_y(\tau)$ characterizes the change of $y$ on the timescale $\tau$~\cite{kartaschoff1978academic}.

The Allan deviation is often used to evaluate the frequency stability of a single clock; however, it can be interpreted as a metric for the stability of synchronization between the reference clock and the measurement target.
The less $\Phi(t)$ alters over time, which means two clocks generate signals of close frequencies or are well synchronized, the lower $\sigma_y(\tau)$ becomes.

We used two rubidium clocks (SIM940, Standard Research Systems) and two Wi-Wi modules.
The rubidium clocks and Wi-Wi modules provide \SI{10}{\mega\hertz} square wave signals.
The relative frequency stability between two signals, or the Allan deviation, was measured by a frequency stability measurement apparatus (SD5M01A, Anritsu).

\begin{figure}[tb]
  \centering
  \includegraphics[width=70mm]{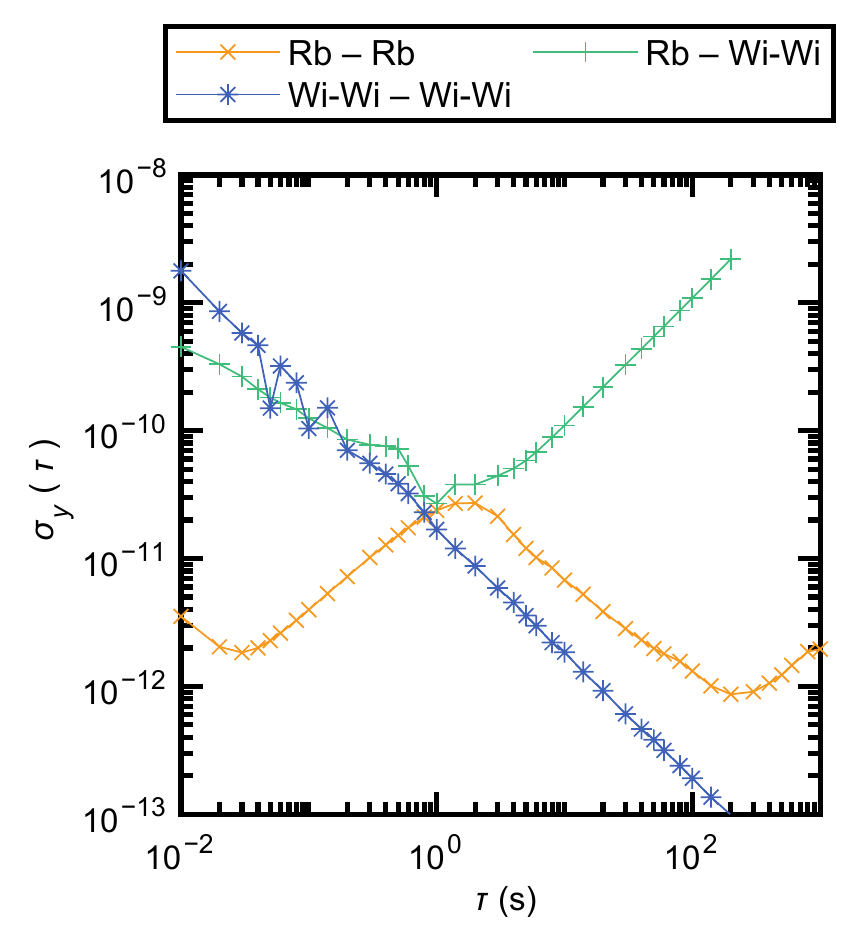}
  \caption{The Allan deviations of fractional frequency offset.
           Lower deviation corresponds to a more stable frequency source.
           A single crystal oscillator in a Wi-Wi module (Rb -- Wi-Wi) was less stable than a rubidium clock (Rb -- Rb).
           However, synchronized Wi-Wi modules (Wi-Wi -- Wi-Wi) outperformed asynchronous rubidium clocks (Rb -- Rb), where $\tau$ was longer than or equal to \SI{1}{s}.
           }
  \label{fig:wi-wi:stability:result}
\end{figure}

The orange $\times$ marks in Fig.~\ref{fig:wi-wi:stability:result} show the Allan deviation when two rubidium clocks were compared.
Similarly, the green $+$ marks indicate the Allan deviation when a rubidium clock and a Wi-Wi module were compared.
Finally, the blue $\ast$ marks show the Allan deviation when two synchronized Wi-Wi modules were compared.

From Fig.~\ref{fig:wi-wi:stability:result}, first, in the timescale $\tau < \SI{1}{\second}$, two rubidium clocks exhibit excellent stability compared with the other two cases.
However, what should be emphasized is that the stability of two Wi-Wi modules outperforms two rubidium clocks when $\tau \geq \SI{1}{\second}$.
Furthermore, the Allan deviation of the Wi-Wi module pair is approximately proportional to $\tau^{-1}$.
This means that additive noise outside the feedback loop was the dominant source of the instability~\cite{kartaschoff1978academic}.
In other words, the result clearly demonstrates that the two-way feedback control of Wi-Wi successfully synchronizes the two oscillators.
In the meantime, the Allan deviation regarding the synchronization between a rubidium clock and a Wi-Wi module, denoted by green $+$ marks, monotonically increases in the timescale beyond approximately $\tau$ of \SI{1}{\second}.
This indicates that a diffusion process is inherent, which prevents synchronization between the atomic clock and the crystal oscillator in Wi-Wi.
This observation is natural since there is no synchronization mechanism between these two.

Additionally, the Allan deviation of two rubidium clocks increases beyond the timescale $\tau$ of approximately \SI{1e2}{\second}.
This indicates that, despite the excellent stability of individual atomic clocks, such individual excellence does not assure the stability of synchronization between the two.

In summing up, the experimental characterization indicates that Wi-Wi excels in synchronization greater than atomic clocks, especially in long timescales.
Once again, what is critical in CSMA/AP is synchronization among multiple devices, which Wi-Wi achieves.

\section{Experimental Demonstration} \label{sec:experiment}

Experimental verification is an essential step in the development of CSMA/AP.
This study experimentally demonstrates CSMA/AP implemented in software-defined radio systems combined with Wi-Wi modules.
Also, we focus on CSMA/AP-T since it encompasses the most foundational aspect of CSMA/AP by exploiting the precise time synchronization of all terminals in the network via Wi-Wi.

To examine the feasibility of CSMA/AP in wireless networks, we conducted three kinds of experiments.
First, we conducted a sensitivity analysis of arbitration point (Sec.~\ref{sec:experiment:ap}) to examine the feasibility of CSMA/AP with the most straightforward setup, while investigating the impacts of CSMA/AP-specific parameters.
Second, we demonstrated CSMA/AP-based communications in a mobile environment (Sec.~\ref{sec:experiment:mobility}), since the capability to deal with moving terminals is one of the significant features in wireless communications compared with wired communications.
Third, we investigated the difference in communication performance of CSMA/AP with or without precise time synchronization by Wi-Wi (Sec.~\ref{sec:experiment:async}) to examine the necessity of time synchronization.

\subsection{Experiment details} \label{sec:experiment:detail}

\subsubsection{Physical Setup}

\begin{figure}[tb]
  \centering
  \begin{minipage}[b]{70mm}
    \centering
    \includegraphics[width=70mm]{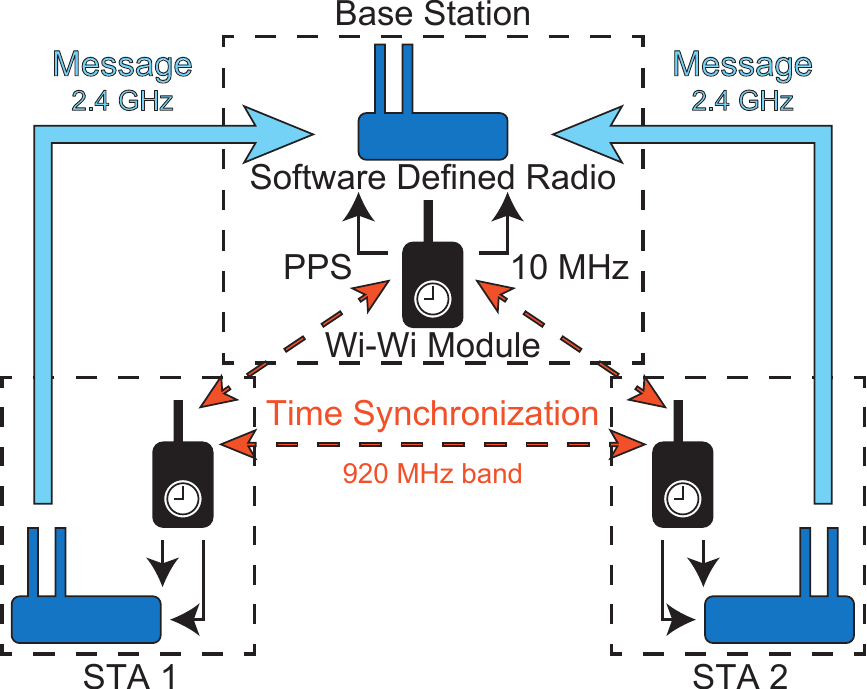}
    \subcaption{}
    \label{fig:experiment:detail:physical:network}
  \end{minipage}
  \begin{minipage}[b]{70mm}
    \centering
    \includegraphics[width=70mm]{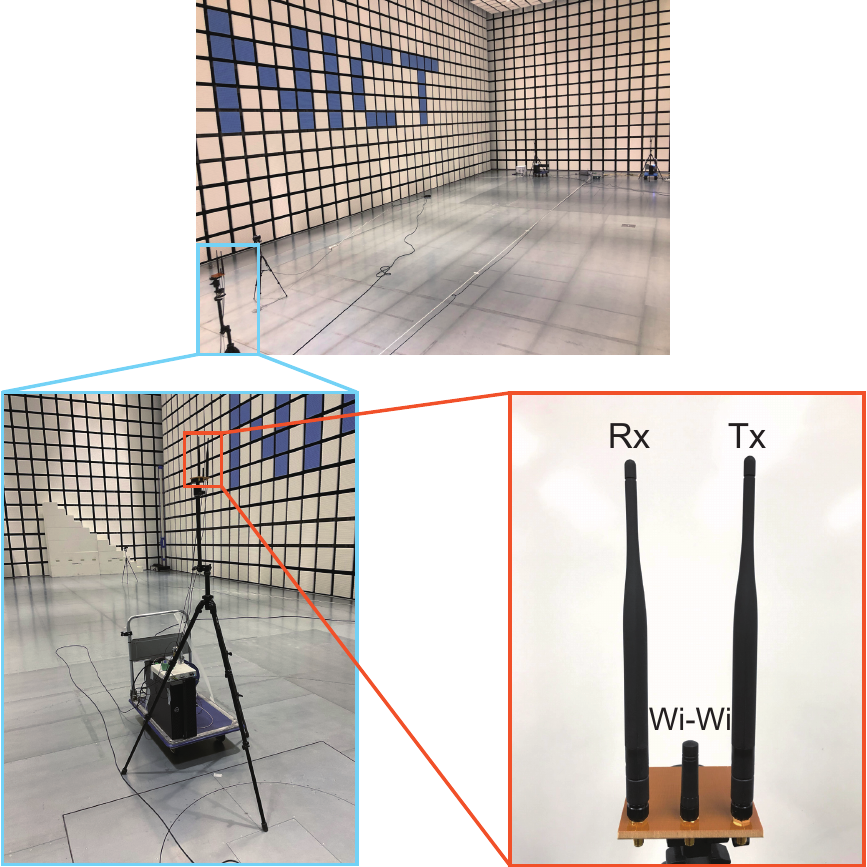}
    \subcaption{}
    \label{fig:experiment:detail:physical:chamber}
  \end{minipage}
  \caption{Design and experimental setup to verify the principle of CSMA/AP-T.
  \subref{fig:experiment:detail:physical:network}~Schematic diagram of the network.
  One of three terminals acts as a base station, and the others, called STA~1 and STA~2, transmit packets alternately.
  Each terminal has a software-defined radio device (SDR) and a Wi-Wi module.
  \subref{fig:experiment:detail:physical:chamber}~Photograph of the electromagnetic anechoic chamber and the apparatuses.
  All faces of the chamber other than the floor were covered with radio-absorbing materials.
  A computer controlled and recorded the behaviors of the paired SDR and Wi-Wi module.
  Antennas were positioned at the top of a tripod, about two meters high.
  Two long antennas were connected to an SDR, and one short antenna was for Wi-Wi module.
  }
  \label{fig:experiment:detail:physical}
\end{figure}

Fig.~\ref{fig:experiment:detail:physical:network} schematically illustrates the architecture of the experimental system where a star-topology wireless network is configured.
One terminal operates as a base station (BS) while the other two work as terminal stations called STA~1 and STA~2.
Here we consider the uplink data transmission from STA~1 to BS and STA~2 to BS.
The data transfer is via a \SI{2.4}{\giga\hertz} carrier by software-defined radio devices (SDR).
Each of the terminals is combined with a Wi-Wi module, leading to precise time synchronization among all terminals.
The Wi-Wi module uses the \SI{920}{\mega\hertz} band for communications and provides a 1 PPS trigger signal and a \SI{10}{\mega\hertz} reference signal to the connected SDR.
Each of the SDRs is connected to a host computer for data logging.

Fig.~\ref{fig:experiment:detail:physical:chamber} shows an overview of the experimental environment and the apparatuses.
All experiments were conducted in an electromagnetic anechoic chamber with five faces, except for the floor which was covered with absorbing materials~\cite{fujii2017nict}.

\subsubsection{Communication Specification}

\begin{figure}
  \centering
  \includegraphics[width=70mm]{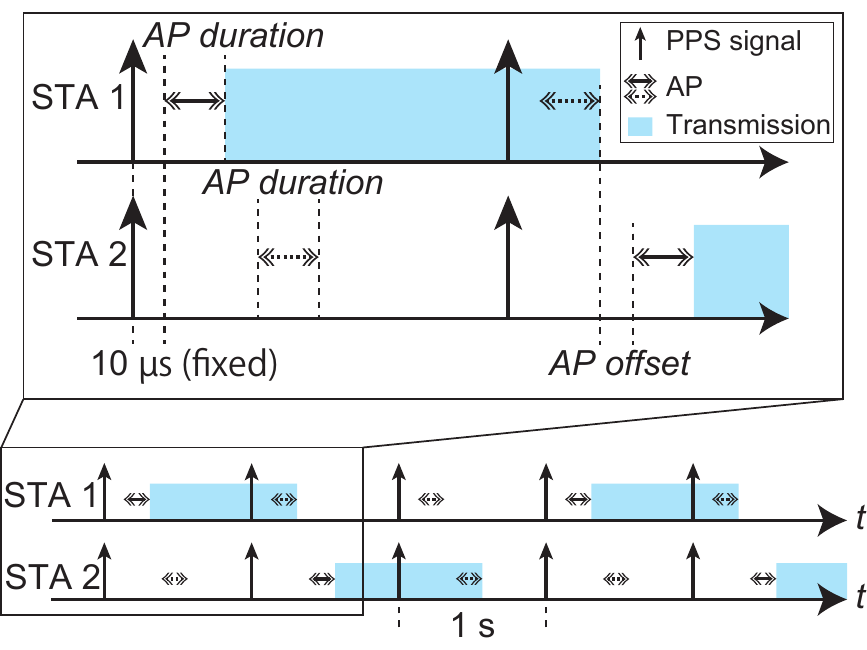}
  \caption{Timing chart of SDRs' behavior.
  Each SDR measures time starting from rising edges of PPS trigger signals and acts according to a predetermined schedule.
  The SDR performs carrier sense during each AP and starts sending a packet after the AP if the channel is not occupied.
  }
  \label{fig:experiment:detail:communication}
\end{figure}

Fig.~\ref{fig:experiment:detail:communication} shows the timing chart of the behavior of the terminals or SDRs which are operated in CSMA/AP.
An SDR waits for a designated trigger signal via the PPS signal provided by the Wi-Wi module, which specifies the starting time of the AP.
The SDR senses the carrier during AP.
Then, if the channel is vacant, the SDR starts transmission.
That is to say, the interval of consecutive APs for STA~1 and STA~2 is \SI{1}{\second}.
In the present experimental system, we used Differential Binary-Phase Shift Keying (DBPSK) as the modulation format, with a symbol rate of \SI{500}{\kilo symbols\per\second}.
The size of a packet was \SI{500}{\kilo\bit s}; therefore, one transmission took \SI{1}{\second}.
The duration of APs and the timing offset of APs, which is examined in detail later in Sec.~\ref{sec:experiment:ap}, were configured with a \SI{1}{\micro\second} resolution.
These specifications and parameter settings originated from technical issues with our current SDR devices.
Technological developments to improve these values are now underway.

STA~1 first transmitted a packet for one second, followed by the transmission from STA~2, and then both STA~1 and STA~2 did not transmit data in the following slot if the system was working properly.
Then the same procedure repeated.
More details about carrier sense and packet contents are described in Appendices~\ref{sec:impl:carrier_sense} and \ref{sec:impl:packet_contents}.

\subsection{Sensitivity Analysis of Arbitration Point} \label{sec:experiment:ap}

The precision timing arrangement among the terminals is critical for CSMA/AP via APs.
Therefore, we first investigate the crucial parameters associated with APs and examine their impact on communications performance.

We consider two parameters: ({\romannumeral 1}) AP duration and ({\romannumeral 2}) AP offset.
AP duration specifies the time length of a single AP, whereas AP offset is the time gap of consecutive APs, as schematically shown in Fig.~\ref{fig:experiment:detail:communication}.
We checked the bit error rate (BER) of the uplink data transfer from STAs to BS as a function of AP duration ranging from \SI{1}{\micro\second} to \SI{4096}{\micro\second} while fixing AP offset to \SI{100}{\micro\second}.
Similarly, AP offset dependency was examined by setting AP offset from \SI{1}{\micro\second} to \SI{1024}{\micro\second}, with AP duration being fixed at \SI{900}{\micro\second}.
In this experiment, the antennas were located at the vertices of an equilateral triangle with a side length of 5 meters.
For each setup, data logging was performed for approximately \SI{5}{\minute}.

\begin{figure}[tb]
  \centering
  \includegraphics[width=70mm]{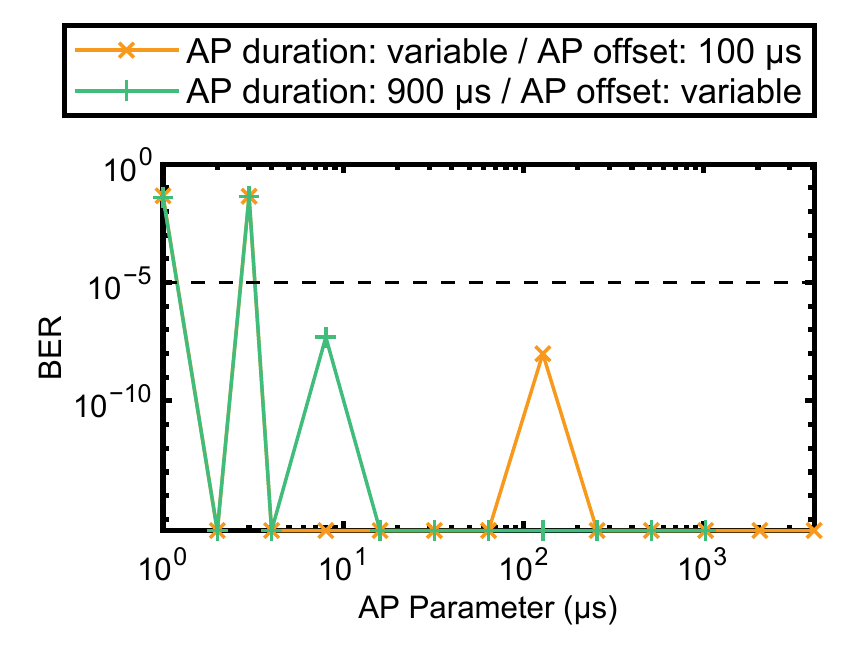}
  \caption{Mean bit error rates (BER) of transferred packets during each experiment.
          The bottom edge of the graph means that BER is equal to 0.
          The black dashed line represents $10^{-5}$, which is the required level of reliability for URLLC~\cite{saad2020network} (note that there was no error-correcting code in these communications).
          }
  \label{fig:experiment:ap:result}
\end{figure}

The $+$ and $\times$ marks in Fig.~\ref{fig:experiment:ap:result} show the mean BER of the data in transferred packets as a function of AP duration and AP offset, respectively.
When both parameters were larger than or equal to \SI{4}{\micro\second}, the BER was below $10^{-5}$, indicating that the communications were successfully accomplished.
Meanwhile, a significant error was observed with parameters being \SI{1}{\micro\second} and \SI{3}{\micro\second}; we speculate that such error may be caused by a technical issue in our current experimental setup of SDRs.
Besides, exploration of much smaller AP parameters will be the topic of one of our future studies, in order to realize a massive number of participants, as reported in~\cite{yamasaki2021access}.
Based on these results, we used 5 or \SI{10}{\micro\second} for AP duration and AP offset in the subsequent experiments.

\subsection{Mobility Demonstration} \label{sec:experiment:mobility}

One significant feature in wireless communication is the capability to deal with mobility.
From the viewpoint of CSMA/AP-T, it is critical to maintain time-synchronization among the terminals, even in mobile environments.

\subsubsection{Distance Dependency}

\begin{figure}[tb]
  \centering
  \includegraphics[width=70mm]{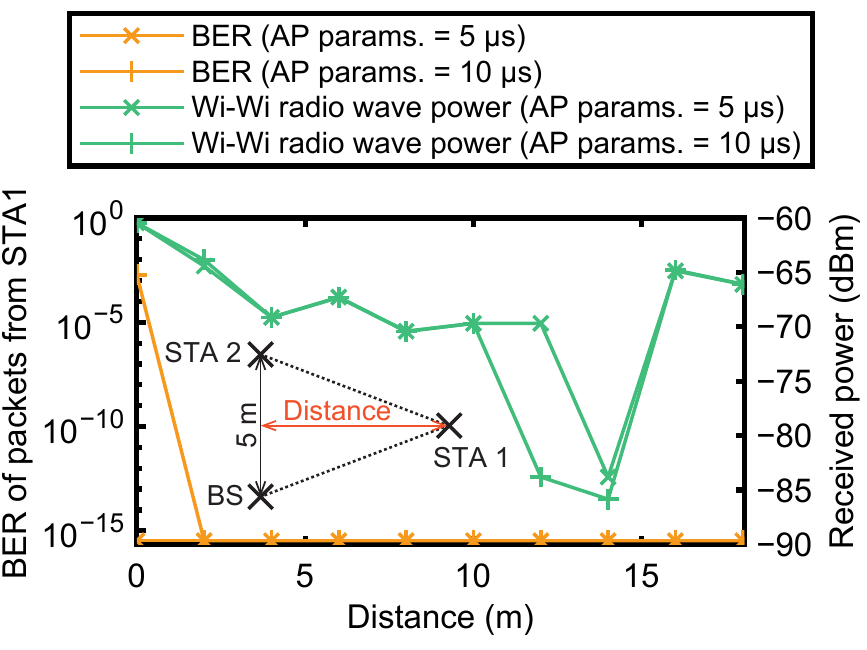}
  \caption{Mean BER of transferred packets and mean radio wave strength at STA~1 as a function of the inter-terminal distance.
          The bottom edge of the vertical axis of BER means that BER is zero, showing that communications were successful in this range of inter-terminal distance.
          Wi-Wi became weaker as the inter-terminal distance increased.
          The power level of Wi-Wi was raised when the distance was farther than \SI{14}{\metre}.
  }
  \label{fig:experiment:mobility:preliminary}
\end{figure}

We first examine the range of successful transmission among terminals to determine the range of the mobile terminal.
The positions of BS, STA~1, and STA~2 formed an isosceles triangle, as shown in the inset of Fig.~\ref{fig:experiment:mobility:preliminary}.
The positions of BS and STA~2 were fixed and were \SI{5}{\meter} apart, whereas the location of STA~1 was configured differently while being equidistant to BS and STA~2.
Precisely, the distance between STA~1 and the baseline connecting BS and STA~2 was set to \SI{0}{\metre}, \SI{2}{\metre}, \ldots, \SI{18}{\metre}.

The orange $\times$ and $+$ marks in Fig.~\ref{fig:experiment:mobility:preliminary} show the BER of transmitted data from STA~1 captured by BS when the parameter setting of AP duration and AP offset parameters were both set to \SI{5}{\micro\second} and both set to \SI{10}{\micro\second}, respectively.
The bottom edge of the vertical axis for BER means that BER was zero, demonstrating that the communication was successful at all distances in the range of inter-terminal distances.

Also, the green marks represent the radio power level of Wi-Wi measured at STA~1, with AP parameter settings being the same as described above.
The power gradually decreased as the distance between STA~1 and BS increased.
Indeed, the power level became too weak when the distance reached \SI{14}{\metre}, resulting in the unsuccessful operation of the system.
Therefore, the operating power of Wi-Wi needed to be increased when the distance was between \SI{14}{\metre} and \SI{18}{\metre}.
Besides, BS could not catch radio waves from STA~1 at the \SI{16}{\metre} position, seemingly due to a null point; horizontally shifting the position of STA~1 by about \SI{15}{\centi\metre} resulted in successful transmission.

\subsubsection{Mobility Demonstrations}

\begin{figure}[tb]
  \centering
  \begin{minipage}{70mm}
    \centering
    \includegraphics[width=70mm]{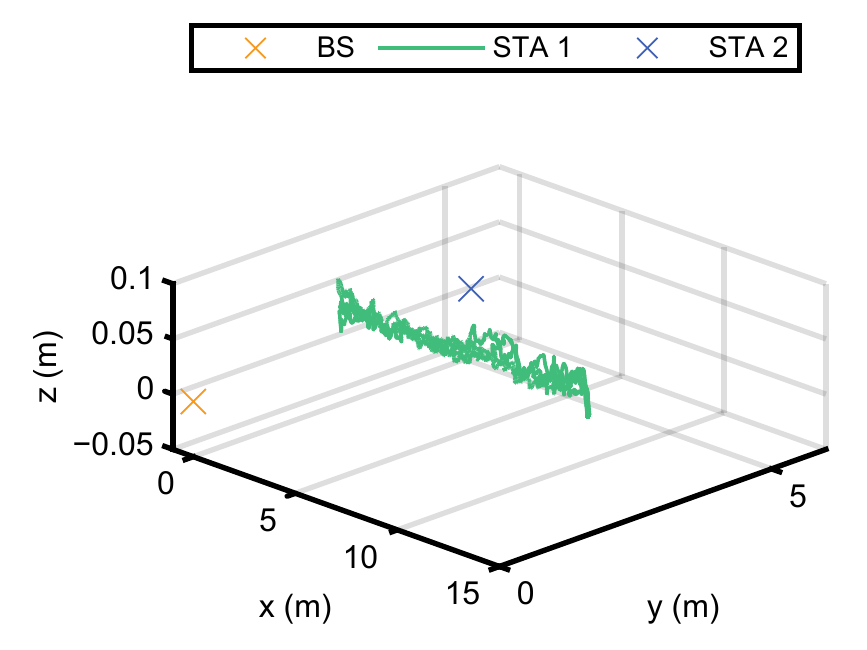}
    \subcaption{}
    \label{fig:experiment:mobility:locus:straight}
  \end{minipage}
  \begin{minipage}{70mm}
    \centering
    \includegraphics[width=70mm]{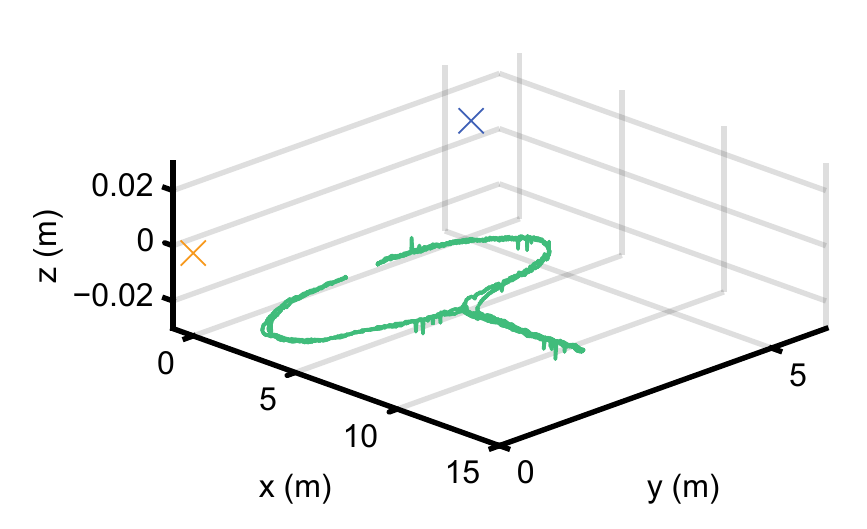}
    \subcaption{}
    \label{fig:experiment:mobility:locus:curved}
  \end{minipage}
  \caption{STA~1's loci tracked with a total station.
  \subref{fig:experiment:mobility:locus:straight}~Linear locus.
  \subref{fig:experiment:mobility:locus:curved}~Curved locus.}
  \label{fig:experiment:mobility:locus}
\end{figure}

Here we show the mobility demonstrations.
The positions of BS and STA~2 were fixed with a \SI{5}{\metre} distance between them, which were the same as in the previous section, as indicated by the orange and blue $\times$ marks in Fig.~\ref{fig:experiment:mobility:locus}, respectively.
STA~1 was the mobile terminal, whose position was tracked by an externally arranged instrument, the so-called total station, which is a coordinate measurement apparatus composed of a laser range finder and an electronic transit theodolite (GT-1201/JBWD, Topcon Positioning Systems).
A 360\degree~reflective prism was attached to STA~1 so that the total station measured and recorded the position of the prism, or STA~1.
More details about position tracking are described in Appendix~\ref{sec:impl:tracking}.
AP duration and AP offset were both \SI{10}{\micro\second}.
Based on the distance limitations revealed in the previous section, the distance between STA~1 and BS was within \SI{14}{\metre} so that the power of Wi-Wi did not have to be reconfigured.
In the mobility demonstrations, we examined two kinds of trajectories of the mobile terminal (STA~1):
one was a straight profile, and the other was a circular one, as shown in Fig.~\ref{fig:experiment:mobility:locus:straight} and \ref{fig:experiment:mobility:locus:curved}, respectively, where the green curves indicate the trace of STA~1 captured by the total station.

\begin{figure}[tb]
  \centering
  \begin{minipage}{70mm}
    \centering
    \includegraphics[width=70mm]{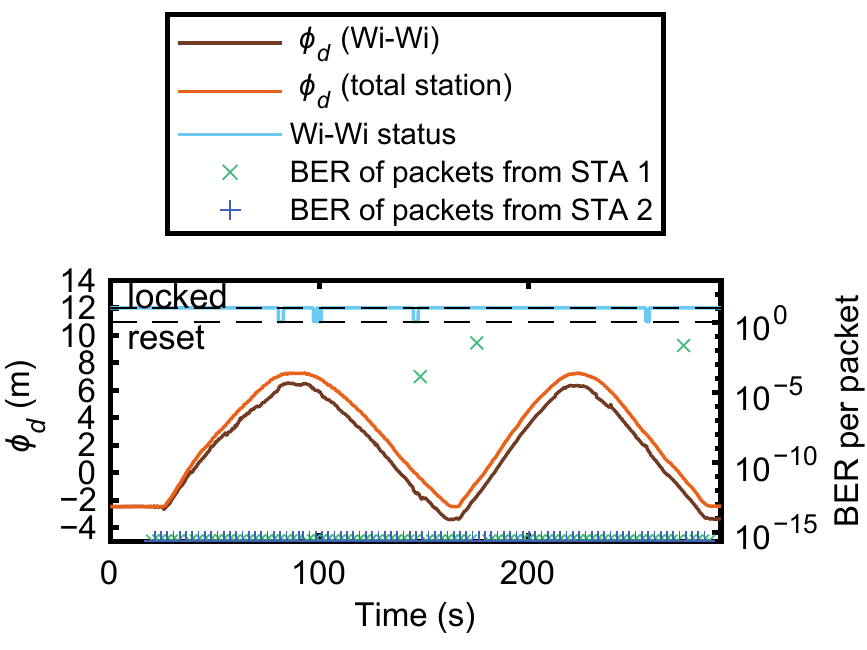}
    \subcaption{}
    \label{fig:experiment:mobility:result:straight}
  \end{minipage}
  \begin{minipage}{70mm}
    \centering
    \includegraphics[width=70mm]{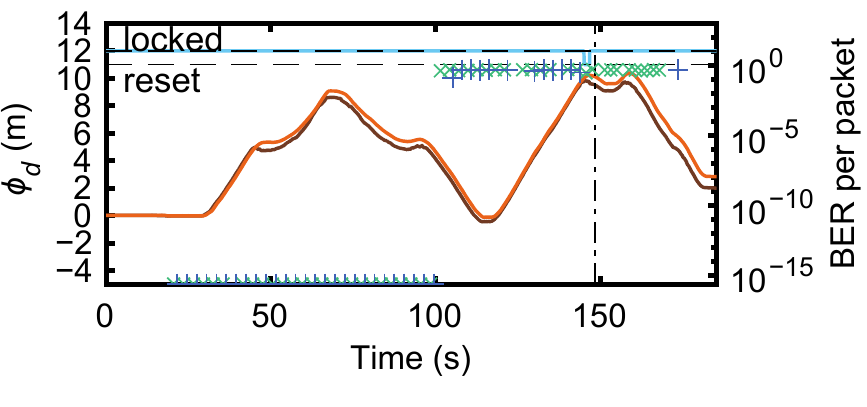}
    \subcaption{}
    \label{fig:experiment:mobility:result:curved}
  \end{minipage}
  \caption{The distance variation $\phi_d$ measured with the total station and Wi-Wi modules, BER of each transferred packet, and the Wi-Wi status during each experiment.
  The bottom edge of the graph for BER represents zero.
  \subref{fig:experiment:mobility:result:straight}~Result of the experiment with the linear locus.
   The communication was successful; however, Wi-Wi modules measured the distance variation with a maximum error of about \SI{1.4}{\metre}.
  \subref{fig:experiment:mobility:result:curved}~Result of the experiment with the curved locus.
  The communication became unsuccessful during this experiment.
  In addition to that, PPS trigger signals became asynchronous at the time indicated with the long dashed-dotted line.}
  \label{fig:experiment:mobility:result}
\end{figure}

The orange curve in Fig.~\ref{fig:experiment:mobility:result:straight} represents the time evolution of the distance between BS and STA~1 measured with the total station, denoted by $\phi_d$ (total station), when STA~1 travels back and forth along a straight line, as shown in Fig.~\ref{fig:experiment:mobility:locus:straight}.
Remember that Wi-Wi measures the inter-device distance by exchanging the time and phase information.
The brown curve in Fig.~\ref{fig:experiment:mobility:result:straight} shows the time evolution of the distance between BS and STA~1 measured by Wi-Wi, denoted by $\phi_d$ (Wi-Wi), which exhibits good agreement with the orange curve obtained by the total station.
This result means that Wi-Wi successfully tracked the distance variation due to the mobility of STA~1.

The green $\times$ and blue $+$ marks in Fig.~\ref{fig:experiment:mobility:result:straight} demonstrate the BER per packet regarding the data transfer from STA~1 to BS and from STA~2 to BS, respectively, where the bottom line means error-free communications were accomplished.
We can observe that there were no errors from STA~2 to BS.
Also, almost completely error-free communications were also realized for the mobile terminal (STA~1), demonstrating successful mobile communications.

Similarly, Fig.~\ref{fig:experiment:mobility:result:curved} summarizes the monitoring results of the evolution of the distance measured by the total station, BER per packet, and the status of Wi-Wi synchronization when STA~1 followed the circular trajectory shown in Fig.~\ref{fig:experiment:mobility:locus:curved}.
The orange and brown curves exhibit similar traces, demonstrating that the distance measurements by Wi-Wi worked well to grasp the ground truth distance profile measured by the total station.
On the other hand, the BER per packet resulted in a higher error rate after approximately \SI{100}{\second}.
Furthermore, we confirmed that the PPS signals from the terminals became asynchronous at approximately \SI{150}{\second}, indicated by the vertical dashed-dot line in Fig.~\ref{fig:experiment:mobility:result:curved}.
The reason behind this phenomenon has not been specified yet, and this will be examined in future studies.

The inter-device distances measured by Wi-Wi modules successfully tracked the mobile terminal in the demonstrations above.
Precisely speaking, however, there were at most approximately \SI{1.4}{\metre} differences between $\phi_d$ measured by the Wi-Wi modules and that by the total station.
We consider that these errors are due to the miscalculation of $M_c$ in \eqref{eq:wi-wi:distance} caused by the reflection of radio waves on the floor.
In the previous research, the distance between antennas was set to about \SI{400}{\milli\metre}~\cite{panta2019sam}, which was much smaller than the distance between terminals in this experiment, and the effect of reflections might have been negligible in that previous experiment.

Finally, we discuss the speed limit of the mobile stations.
The relative speed of terminals is limited to $\lambda / 4T$ based on the principle of Wi-Wi, where $\lambda$ is the wavelength of the carrier used in Wi-Wi communications and $T$ is the interval of successive Wi-Wi data revision at the terminal.
Currently, $\lambda$ is approximately \SI{320}{\milli\metre}, and $T$ is \SI{50}{\milli\second}; therefore, the upper bound of the speed is about \SI{6}{\kilo\metre\per\hour}, which is the speed of a slowly walking pedestrian.
The mobility of the station (STA~1) was controlled by humans who walked at an average walking pace, which satisfies the above bound.

\subsection{CSMA/AP in Asynchronous Situations} \label{sec:experiment:async}

Precise time synchronization is indispensable in CSMA/AP.
Here we experimentally examine the resultant communication with and without Wi-Wi-based time synchronization.
The configuration of the terminals was the same as in Sec.~\ref{sec:experiment:ap}; the three stations were located \SI{5}{\metre} apart from each other.
AP duration and AP offset were both set to \SI{5}{\micro\second}.

The status of a Wi-Wi module can be configured as ``synchronized'' or ``desynchronized.''
Data logging was conducted for approximately \SI{5}{\minute} with the Wi-Wi modules synchronized and for \SI{20}{\minute} when they were desynchronized.

In the meantime, the time difference between PPS signals provided by STA~1 and STA~2 were monitored by an oscilloscope.
Remember that PPS signals specify the timing of APs at each terminal, meaning that the time difference between them is critical in ensuring collision-free communications.

\begin{figure}[tb]
  \centering
  \begin{minipage}{70mm}
    \centering
    \includegraphics[width=70mm]{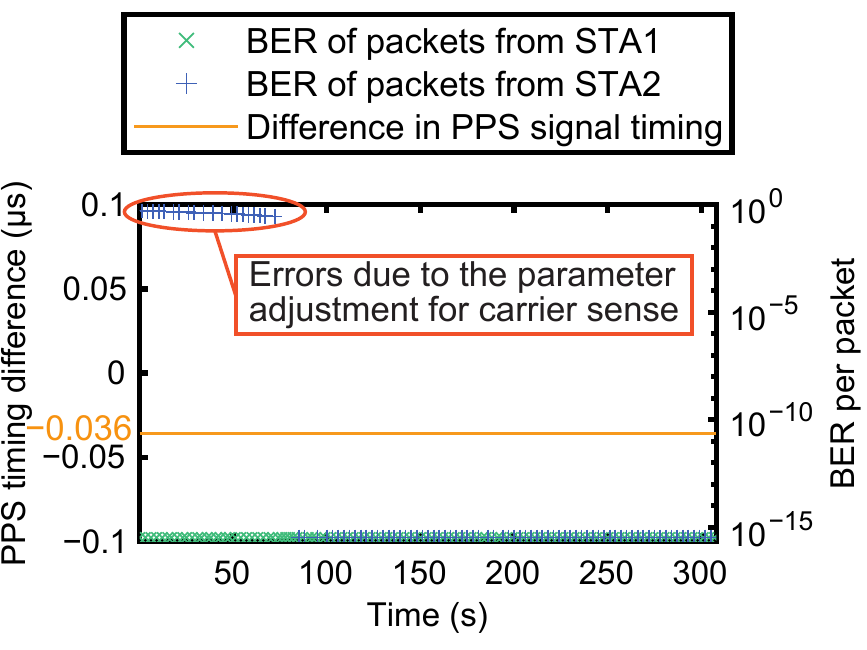}
    \subcaption{}
    \label{fig:experiment:status:result:synced}
  \end{minipage}
  \begin{minipage}{70mm}
    \centering
    \includegraphics[width=70mm]{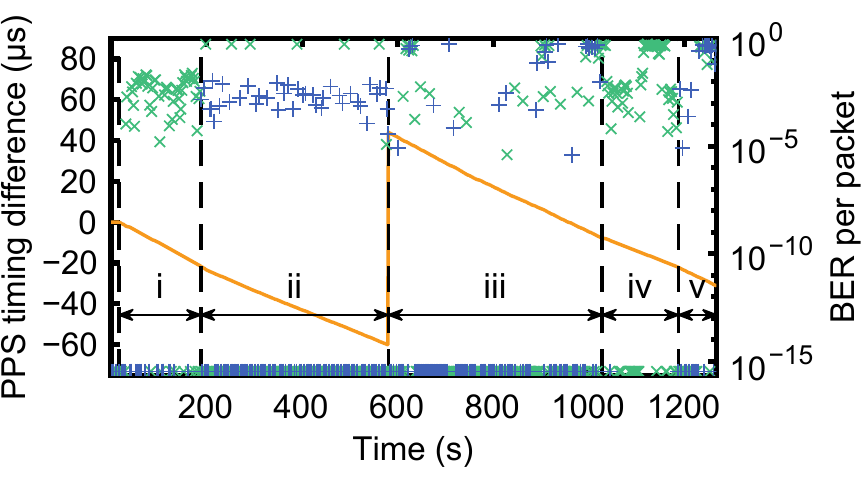}
    \subcaption{}
    \label{fig:experiment:status:result:desynced}
  \end{minipage}
  \caption{The difference in the timing of PPS trigger signals for STA~1 and STA~2 and BER of each transferred packet.
  \subref{fig:experiment:status:result:synced}~When Wi-Wi modules were synchronized.
  The difference in PPS signal timing did not change at a detectable level during this experiment.
  Communication errors at the beginning of this experiment were due to the parameter adjustment for carrier sense.
  \subref{fig:experiment:status:result:desynced}~When Wi-Wi modules were asynchronous.
  Each time region ({\romannumeral 1} -- {\romannumeral 5}) is characterized by a different communication status, summarized in Table~\ref{table:experiment:status:result}.
  }
  \label{fig:experiment:status:result}
\end{figure}

Fig.~\ref{fig:experiment:status:result:synced} and \ref{fig:experiment:status:result:desynced} represent the time evolution of the BER per packet measured by the BS transmitted from STA~1 and STA~2, respectively.
Note that the bottom limit of the vertical axis of BER means zero error.
The orange curve therein shows the timing difference between STA~1 and STA~2.
Here the time difference takes a positive value when the PPS signals from STA~1 arrive earlier than those from STA~2.
Fig.~\ref{fig:experiment:status:result:synced} and \ref{fig:experiment:status:result:desynced} show synchronized and desynchronized situations, respectively.

In Fig.~\ref{fig:experiment:status:result:synced}, with Wi-Wi being synchronized, the BER per packet from STA~1 and STA~2 both exhibited no errors after approximately \SI{70}{\second}.
An automatic adjustment was in operation for the carrier sense signal level in the initial durations.
The details of carrier sense are described in Appendix~\ref{sec:impl:carrier_sense}.
Once it was completed, zero BER was realized.
Furthermore, the PPS timing difference was stable and almost zero throughout the measurement.

Conversely, in Fig.~\ref{fig:experiment:status:result:desynced}, significant packet transmission errors occurred.
In particular, some packets from STA~2 were not recognized by BS in some of the time regions.
Furthermore, the time difference of PPS signals was monotonically changing over time due to specific diffusion mechanisms inherent in the oscillators.
As examined by the following, there are some different kinds of data transfer in the desynchronized situations.
Table~\ref{table:experiment:status:result} summarizes the communications status during each time region ({\romannumeral 1} -- {\romannumeral 5}) in Fig.~\ref{fig:experiment:status:result:desynced}.

\begin{table}[tb]
    \centering
    \caption{
    The communication status during each time region ({\romannumeral 1} -- {\romannumeral 5}) in Fig.~\ref{fig:experiment:status:result:desynced}.The items shown in italic indicate that the status is different from the intended ideal case. The items denoted with underlining show that collisions occurred.
    }
    \label{table:experiment:status:result}
    \begin{tabular}{crclc} \toprule
         \textbf{Time region} & \multicolumn{3}{c}{\textbf{PPS timing difference}} & \textbf{Transmission order} \\ \midrule
         (Ideal case)& \multicolumn{3}{c}{$\ll \SI{1}{\micro\second}$} & STA~1 $\rightarrow$ STA~2 $\rightarrow$ vacant \\ \midrule
         {\romannumeral 1} & \SI{0.0}{\micro\second}   & -- & \textit{\SI{-21.3}{\micro\second}} & \underline{\textit{Simultaneous}} \\
         {\romannumeral 2} & \textit{\SI{-21.6}{\micro\second}} & -- & \textit{\SI{-60.1}{\micro\second}} & \textit{STA~2 $\rightarrow$ STA~1 $\rightarrow$ vacant} \\
         {\romannumeral 3} & \textit{\SI{44.1}{\micro\second}}  & -- & \textit{\SI{-7.5}{\micro\second}}  & STA~1 $\rightarrow$ STA~2 $\rightarrow$ vacant \\
         {\romannumeral 4} & \textit{\SI{-7.7}{\micro\second}}  & -- & \textit{\SI{-21.8}{\micro\second}} & \underline{\textit{Simultaneous}} \\
         {\romannumeral 5} & \textit{\SI{-22.0}{\micro\second}} & -- & \textit{\SI{-31.3}{\micro\second}} & \textit{STA~2 $\rightarrow$ STA~1 $\rightarrow$ vacant} \\ \bottomrule
    \end{tabular}
\end{table}

\begin{figure}[tb]
  \centering
  \begin{minipage}{70mm}
    \centering
    \includegraphics[width=70mm]{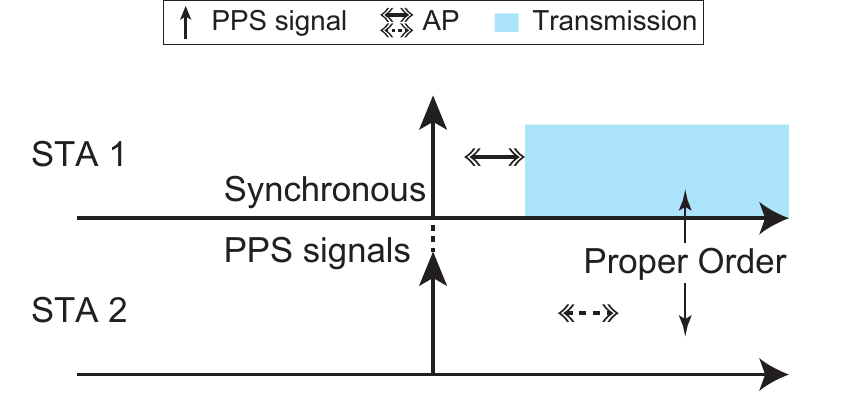}
    \subcaption{}
    \label{fig:experiment:status:timing:indicated}
  \end{minipage}
  \begin{minipage}{70mm}
    \centering
    \includegraphics[width=70mm]{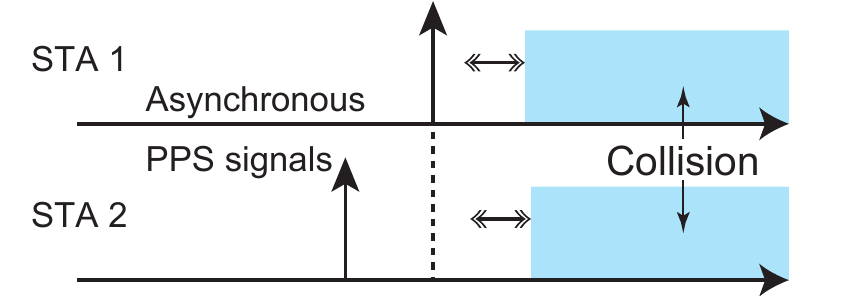}
    \subcaption{}
    \label{fig:experiment:status:timing:interference}
  \end{minipage}
  \begin{minipage}{70mm}
    \centering
    \includegraphics[width=70mm]{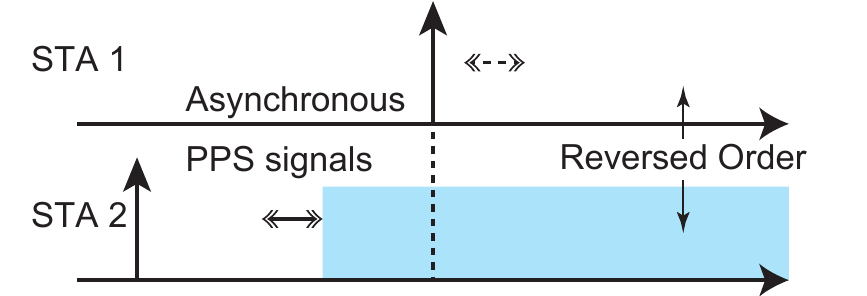}
    \subcaption{}
    \label{fig:experiment:status:timing:reverse}
  \end{minipage}
  \caption{Timing chart of APs and packet transmissions over each time region in Fig.~\ref{fig:experiment:status:result:desynced}.
  \subref{fig:experiment:status:timing:indicated}~Indicated timing. APs are correctly positioned without overlapping, and no collision occurs.
  \subref{fig:experiment:status:timing:interference}~APs can overlap when PPS signals arrive asynchronously at STA~1 and STA~2, and STA~1 fails to detect transmissions from STA~2.
  In this case, packets are simultaneously transmitted from STA~1 and STA~2, and collisions occur.
  \subref{fig:experiment:status:timing:reverse}~If PPS signals arrive too early at STA~2 than STA~1, the transmission order switches.
  In this case, collisions do not occur; however, this switching can cause other collisions when there are more APs in one cycle.
  }
  \label{fig:experiment:status:timing}
\end{figure}

Fig.~\ref{fig:experiment:status:timing} summarizes the proper and improper operations of STA~1 and STA~2 in the form schematic timing chart.
First, Fig.~\ref{fig:experiment:status:timing:indicated} represents the proper situation where STA~1 and STA~2 are successfully synchronized where the PPS signals denoted by vertical arrows are completely time aligned.
The time duration denoted by horizontal arrows shows the AP duration.
In this situation, STA~1 starts data transmission after the AP of STA~1, indicated by the blue region.
STA~2 also goes through the AP durations after the time gap specified by AP offset.
STA~2 senses carriers in its AP durations, and STA~2 postpones data transmission, leading to no collision.

Conversely, in Fig.~\ref{fig:experiment:status:timing:interference}, STA~1 and STA~2 become asynchronous where the PPS signal of STA~2 is earlier than STA~1.
In this situation, AP for STA~1 and that for STA~2 overlap, leading to simultaneous data transmission by both STAs.
Therefore, collisions occur.
In Fig.~\ref{fig:experiment:status:result:desynced}, the time regions marked by {\romannumeral 1} and {\romannumeral 4} were such situations.

Furthermore, Fig.~\ref{fig:experiment:status:timing:reverse} illustrates a situation when the degree of synchronicity becomes more severe.
The PPS signal of STA~2 is way earlier than that of STA~1, which allows STA~2 to send information.
Consequently, STA~1 does not start data transmission.
That is, the order is reversed.
Such a situation is good in terms of avoiding data transmission conflicts.
However, this is not the intended order of the original design.
In view of an increased number of terminals, such an unexpected out-of-order transmission must be avoided to preserve the delay-bounded property.
In Fig.~\ref{fig:experiment:status:result:desynced}, such reversed situations happened in the time regions {\romannumeral 2} and {\romannumeral 5}.

Moreover, even under the proper situation in Fig.~\ref{fig:experiment:status:timing:indicated}, meaning that there are no collisions and no reverse order, it should be emphasized that the BER was significantly high, as observed in the time region {\romannumeral 3} in Fig.~\ref{fig:experiment:status:result:desynced}.
This is because of the difference in frequencies of \SI{10}{\mega\hertz} reference signals due to the asynchronicity among modules.
Transmitters and receivers generate \SI{2.4}{\giga\hertz} signals from those \SI{10}{\mega\hertz} reference signals, and differences in frequencies of reference signals cause the carrier frequency offset.
This induced errors in decoding received signals.

\section{Conclusion} \label{sec:conclusion}

This paper reports successful experimental demonstrations of the principles of Carrier Sense Multiple Access with Arbitration Point (CSMA/AP), which strictly ensures the upper bound of delay, based on software-defined radio combined with Wireless Two-Way Interferometry (Wi-Wi) devices.
This study is the first experimental demonstration of the arbitration of wireless communications using Wi-Wi.
A star-topology network was constructed using a \SI{2.4}{\giga\hertz} band for data communications and a \SI{920}{\mega\hertz} band for time synchronization of Wi-Wi.
The communication by CSMA/AP was performed without bit errors when the positions of the terminals were fixed.
CSMA/AP was also successfully operated when the position of the terminal was dynamically reconfigured, or when the terminal was mobile.
Also, a comparison with desynchronized situations confirmed the significance of precise time synchronization.
The synchronization ability between Wi-Wi-based devices and atomic clocks was also shown by experimentally examining the Allan deviation to emphasize the differences between the notion of frequency stability and synchronization.
There were several factors that affected the performance of CSMA/AP.
The most significant factor was time and frequency synchronization.
This is the fundamental of CSMA/AP, and we throughly investigated its impact with two experiments.
Another factor is the reflection of radio waves.
The reflection of radio waves on the floor disturbed the function of Wi-Wi modules, causing insufficient synchronization and distance measurement.

Although the fundamental principles of CSMA/AP are validated in the present study, there are a variety of remaining issues to be solved.
For example, the packet rate is determined directly by the PPS signal supplied by the Wi-Wi module due to technical issues.
The increased packet rate is an immediate future technological target, which is also essential to exploit the ability to accommodate a massive amount of terminals in CSMA/AP.
Also, what is demonstrated in this study is CSMA/AP-T.
Another variant of CSMA/AP is CSMA/AP-TS, where the inter-terminal distance information is fully utilized.
Experimental examinations of CSMA/AP-TS are also needed.
Autonomous user addition and deletion will be another fundamental issue.
A variety of technical improvements, such as increased data rate and device integration, will be possible on the basis of recent advancements in software-defined radio and related technologies.

In summary, this study paves the way for future communications to benefit from precision time- and space- synchronization for reliable information and communications.

\appendix
\section{Carrier Sense and Arbitration} \label{sec:impl:carrier_sense}

We used Algorithm~\ref{alg:impl:carrier_sense} for carrier sense and arbitration.
There are two hyperparameters: $N$ and $a$.
$N$ is the length of queues $i_\mathrm{rec}$ and $q_\mathrm{rec}$ recording accumulated in-phase and quadrature signals, respectively, and $a$ is the scale factor used when updating thresholds.
We set $N$ to 3 and $a$ to 5 during the mobility demonstration, and $N$ to 10 and $a$ to 2 during the other experiments.
Here, $i_0$ and $q_0$ are the direct-current components of in-phase and quadrature signals, respectively; $i_\mathrm{thresh}$ and $q_\mathrm{thresh}$ are the thresholds for accumulated in-phase and quadrature signals, respectively; $i(t)$ and $q(t)$ are the in-phase and quadrature signals, respectively; and $t_s$ and $t_e$ are the starting and ending time of an AP, respectively.

\begin{figure}[tb]
  \begin{algorithm}[H]
    \caption{Carrier sense and arbitration}
    \label{alg:impl:carrier_sense}
    \begin{algorithmic}[1]
      \State $i_\mathrm{rec} = (i_\mathrm{rec}^1, i_\mathrm{rec}^2, \ldots, i_\mathrm{rec}^N) \gets (0, \ldots, 0)$
      \State $q_\mathrm{rec} = (q_\mathrm{rec}^1, q_\mathrm{rec}^2, \ldots, q_\mathrm{rec}^N) \gets (0, \ldots, 0)$
      \State Set $i_0, q_0, i_\mathrm{thresh}$ and $q_\mathrm{thresh}$
      \Loop
        \State $i_\mathrm{acc} \gets \int_{t_s}^{t_e} (i(t) - i_0) \dd{t}$
        \State $q_\mathrm{acc} \gets \int_{t_s}^{t_e} (q(t) - q_0) \dd{t}$
        \If{$i_\mathrm{acc}> i_\mathrm{thresh}$ and $q_\mathrm{acc} > q_\mathrm{thresh}$}
        \Statex \Comment{The channel is vacant}
          \State $i_\mathrm{rec} \gets (i_\mathrm{rec}^2, \ldots, i_\mathrm{rec}^N, i_\mathrm{acc})$
          \State $q_\mathrm{rec} \gets (q_\mathrm{rec}^2, \ldots, q_\mathrm{rec}^N, q_\mathrm{acc})$
          \If{$N - 1$ or more packets have been sent}
            \State $i_\mathrm{thresh} \gets \frac{a}{N} \sum_{k = 1}^N i_\mathrm{rec}^k$
            \State $q_\mathrm{thresh} \gets \frac{a}{N} \sum_{k = 1}^N q_\mathrm{rec}^k$
          \EndIf
          \State Start transmission
        \Else \Comment{The channel is occupied}
          \State Wait until the next AP
        \EndIf
      \EndLoop
    \end{algorithmic}
  \end{algorithm}
\end{figure}

Adaptive physical carrier sense threshold has been widely investigated~\cite{thorpe2014csto}.
Also, Algorithm~\ref{alg:impl:carrier_sense} follows the idea of CSMA/AP-T examined in~\cite{yamasaki2021access}.
That is, Algorithm~\ref{alg:impl:carrier_sense} is a composition of CSMA/AP-T and dynamic adjustment of carrier sense level, which was developed originally in the present study to make the access control adaptable to various conditions in the experiments.

\section{Packet Contents} \label{sec:impl:packet_contents}

As mentioned in Sec.~\ref{sec:experiment}, each transmitted packet consists of \SI{500}{\kilo\bit s}.
The beginning \SI{113}{\bit s} are the header part, and the following \SI{499887}{\bit s} are the body part.
Table~\ref{table:impl:packet} shows the contents of one packet.

\begin{table}[tb]
  \centering
  \caption{Contents of a packet.
   ``Guard bit,'' ``sync numbers,'' ``header,'' ``body length,'' ``destination,'' and ``body'' are fixed, while ``source'' and ``packet id'' vary.
   ``Source'' is set to 1 when the packet is from STA~1, or 2 otherwise.
   ``Packet id'' is a counter of packets and increases by 1 when a new packet is transmitted from each terminal.}
  \label{table:impl:packet}
  \begin{small}
  \begin{tabular}{cccc} \hline
    \textbf{Item} & Guard bit & Sync number & Header \\
    \textbf{Length (bit)} & 1 & 16 & 16 \\
    \textbf{Content} & 1 & 0xE98A & 0xFFAA \\ \hline
    & & & \\ \hline
    \textbf{Item} & Body length & Source & Destination \\
    \textbf{Length (bit)} & 32 & 16 & 16 \\
    \textbf{Content} & 499887 & vary & 3 \\ \hline
    & & & \\ \hline
    \textbf{Item} & Packet id & Body \\
    \textbf{Length (bit)} & 16 & 499887 \\
    \textbf{Content} &  vary & fixed pattern \\ \hline
  \end{tabular}
  \end{small}
\end{table}

\section{Position Tracking of the Mobile Terminal} \label{sec:impl:tracking}
A total station is an instrument for land surveying, which is a combination of a laser range finder and an electronic theodolite.
It can automatically track a reflective marker (we used a 360\degree~reflective prism) and record the position of the marker.

\begin{figure}
    \centering
    \includegraphics[width=70mm]{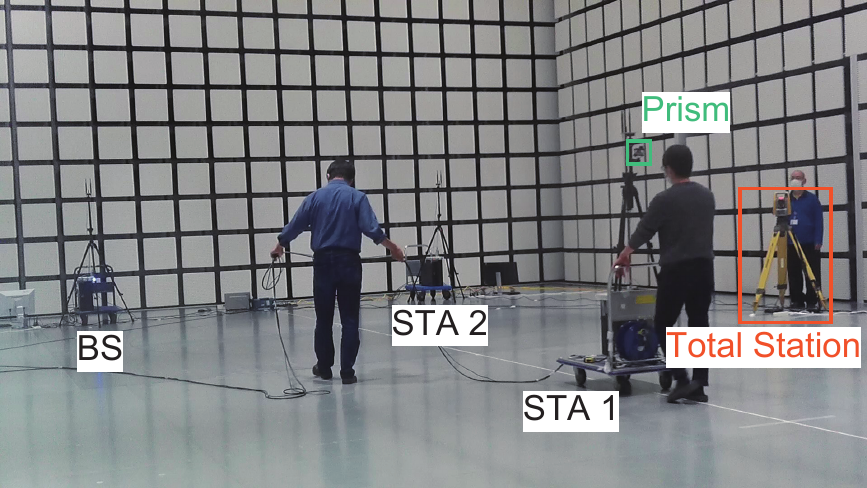}
    \caption{Positions of BS, STA~1, STA~2, the total station, and the prism on STA~1 during the mobility demonstration in the experiment.}
    \label{fig:impl:tracking:position}
\end{figure}

During the experiments, the total station was positioned at the place shown in Fig.~\ref{fig:impl:tracking:position}.
Also, the reflective prism was fixed on the frame of the tripod where STA~1 was attached.
Prior to the experiments, the positional displacement between the prism and the antennas was measured.
The offset was compensated in the experimental data processing step.

\section*{Acknowledgment}

H. Tanaka thanks A. R\"{o}hm from The University of Tokyo for fruitful discussions and for reviewing a draft of this paper.

\bibliographystyle{IEEEtran}
\input{access.bbl}

\end{document}